\documentclass[manuscript]{aastex}

\shorttitle{CME statistics}
\shortauthors{Bein et al.}

\usepackage{natbib}
\usepackage{graphicx}
\usepackage{amsmath}

\begin{document}
\title{Impulsive acceleration of coronal mass ejections: \\ I. Statistics and CME source region characteristics}

\author{B. M. Bein\altaffilmark{1}, S. Berkebile-Stoiser\altaffilmark{1}, A. M. Veronig\altaffilmark{1}, M. Temmer\altaffilmark{1},  \\ N. Muhr\altaffilmark{1}, I. Kienreich\altaffilmark{1}, D. Utz\altaffilmark{1} }
\affil{IGAM/Institute of Physics, University of Graz, Universit{\"a}tsplatz 5, 8010 Graz, Austria}

\and

\author{B. Vr{\v s}nak\altaffilmark{2}}
\affil{Hvar Observatory, Faculty of Geodesy, University of Zagreb, Ka{\v c}i\'{c}eva 26, HR-10000 Zagreb, Croatia}

\date{Received/Accepted}

\begin{abstract}
We use high time cadence images acquired by the STEREO EUVI and COR instruments to study the evolution of coronal mass ejections (CMEs), from their initiation, through the impulsive acceleration to the propagation phase. For a set of 95 CMEs we derived detailed height, velocity and acceleration profiles and statistically analysed characteristic CME parameters: peak acceleration, peak velocity, acceleration duration, initiation height, height at peak velocity, height at peak acceleration and size of the CME source region. The CME peak accelerations derived range from 20 to 6800~m~s$^{-2}$ and are inversely correlated to the acceleration duration and to the height at peak acceleration. 74\% of the events reach their peak acceleration at heights below 0.5~$R_{\odot}$. CMEs which originate from compact sources low in the corona are more impulsive and reach higher peak accelerations at smaller heights. These findings can be explained by the Lorentz force, which drives the CME accelerations and decreases with height and CME size.
\end{abstract}


\maketitle

\section{Introduction}

Coronal mass ejections (CMEs) are sporadic ejections of magnetized plasma from the Sun with masses in the order of 10$^{13}$--10$^{16}$~g \citep{vourlidas2010} and velocities in the range of $\sim$100--3000~km~s$^{-1}$ \citep[e.g.][]{yashiro2004, gopalswamy2009}. They are accelerated by magnetic forces in the solar corona and move outwards into the interplanetary space, where they may severely influence the space weather near the Earth. 

Several case studies of the CME kinematics \citep{gallagher2003, zhang2001, zhang2004, maricic2004, temmer2008, temmer2010} showed that CMEs typically undergo three phases: a gradual evolution, fast acceleration and propagation phase \citep{zhang2001}. In the gradual phase, the CME leading edge rises slowly with velocities of some ten km~s$^{-1}$. At a certain height, the CME undergoes a strong acceleration. How fast and how long the acceleration takes place varies from event to event. After the main impulsive acceleration phase the CME propagates at almost constant velocity or shows a gradual acceleration/deceleration due to the interaction with the ambient solar wind flow during the propagation in the interplanetary space \citep[e.g.][]{gopalswamy2000}.  

Recent studies suggest that the main acceleration of impulsive CMEs occurs at low coronal heights, which are not observable in traditional white light coronographic images \citep[][]{gallagher2003, temmer2008, temmer2010}. Thus, if we are interested in the origin of CMEs and their initial acceleration, we have to observe them from their initiation site close to the solar surface. To observe fast CME accelerations it is important to have image sequences with high time cadence during this important dynamical phase of the CMEs. Since fast and impulsive events severely contribute to our space weather, they are particularly relevant to study.

Acceleration measurements of the early phase of the CME propagation for a statistical sample of 50 events were done by \citet{zhang2006}, who used coronagraphic observations from 1.1~$R_{\odot}$ to 30~$R_{\odot}$ (LASCO C1, C2 and C3), before the LASCO C1 coronagraph failed in 1998 due to the communication loss with the SOHO spacecraft. \citet{vrsnak2007} combined EUV images (SOHO EIT) with coronagraphic observations (MLSO Mark-IV K, LASCO C2 and C3) in order to track CMEs from their initiation site up to about 30~$R_{\odot}$. They analysed a sample of 22 events, which contain predominantly gradual CMEs.  

For our study we used Solar Terrestrial Relations Observatory \citep[STEREO;][]{kaiser2008} data, which provide high time cadence EUV imaging and coronagraphic observations up to 15~$R_{\odot}$ with an overlapping field-of-view (FOV). Based on this data set we derived detailed CME kinematics and acceleration profiles for a sample of 95 impulsive CME events, which occurred during January 2007 and May 2010, representing the largest data set for the study of impulsive CME acceleration so far. In this paper statistics and correlation analysis of the kinematical and dynamical CME characteristics are presented. In our parameter study we focus on the CME peak velocity, peak acceleration, acceleration duration, height at peak velocity, height at peak acceleration, initiation height and the CME source region size. The relation of the CME characteristics to the associated flare, filament eruption, large scale EUV waves and magnetic topology will be presented in a forthcoming paper.

\section{Data and data reduction}

\begin{figure*}
	\centering
		\includegraphics[scale=0.81]{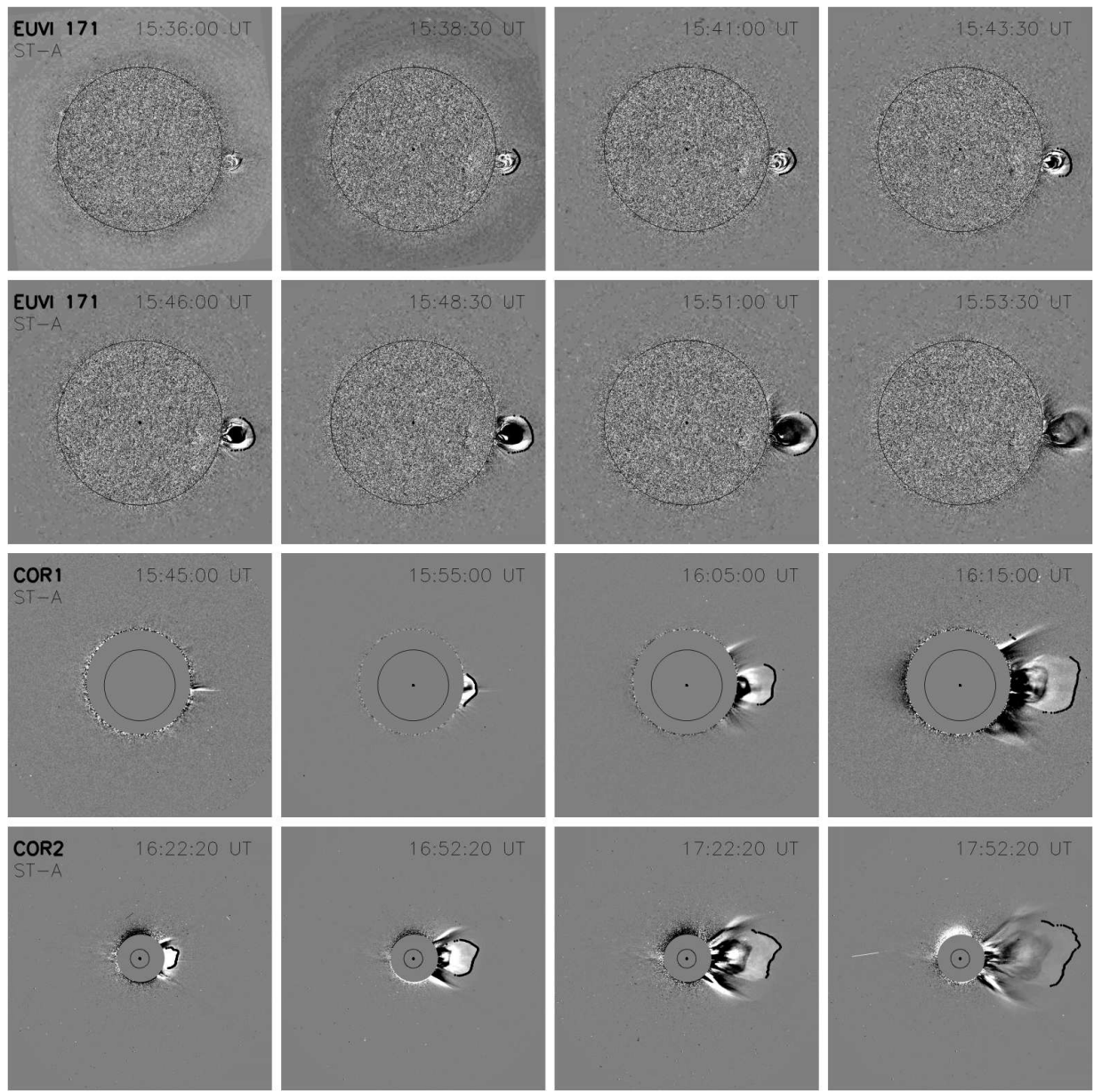}
	\caption{Time sequence of EUVI 171~{\AA} (first and second row), COR1 (third row) and COR2 (bottom row) images of the CME observed on 5 April 2008 by STEREO-A. The curved line marks the identified CME leading edge, the straight line indicates the propagation direction. The CME kinematics derived for this event is plotted in Figs.~\ref{plotheight} and \ref{plothva}. All EUVI images are plotted in an $x$-range of $[-1600'', +1600'']$ and in an $y$-range of $[-1470'', +1730'']$, the selected COR1 range for both directions is
	$[-3720'', +3720'']$ and the COR2 range $[-14\,000'',+14\,000'']$. 
	Note that not all images available are shown. A movie for the whole event including all images available is included in the electronic supplement.}
	\label{sequence}
\end{figure*}

\begin{figure*}
	\centering
		\includegraphics[scale=0.81]{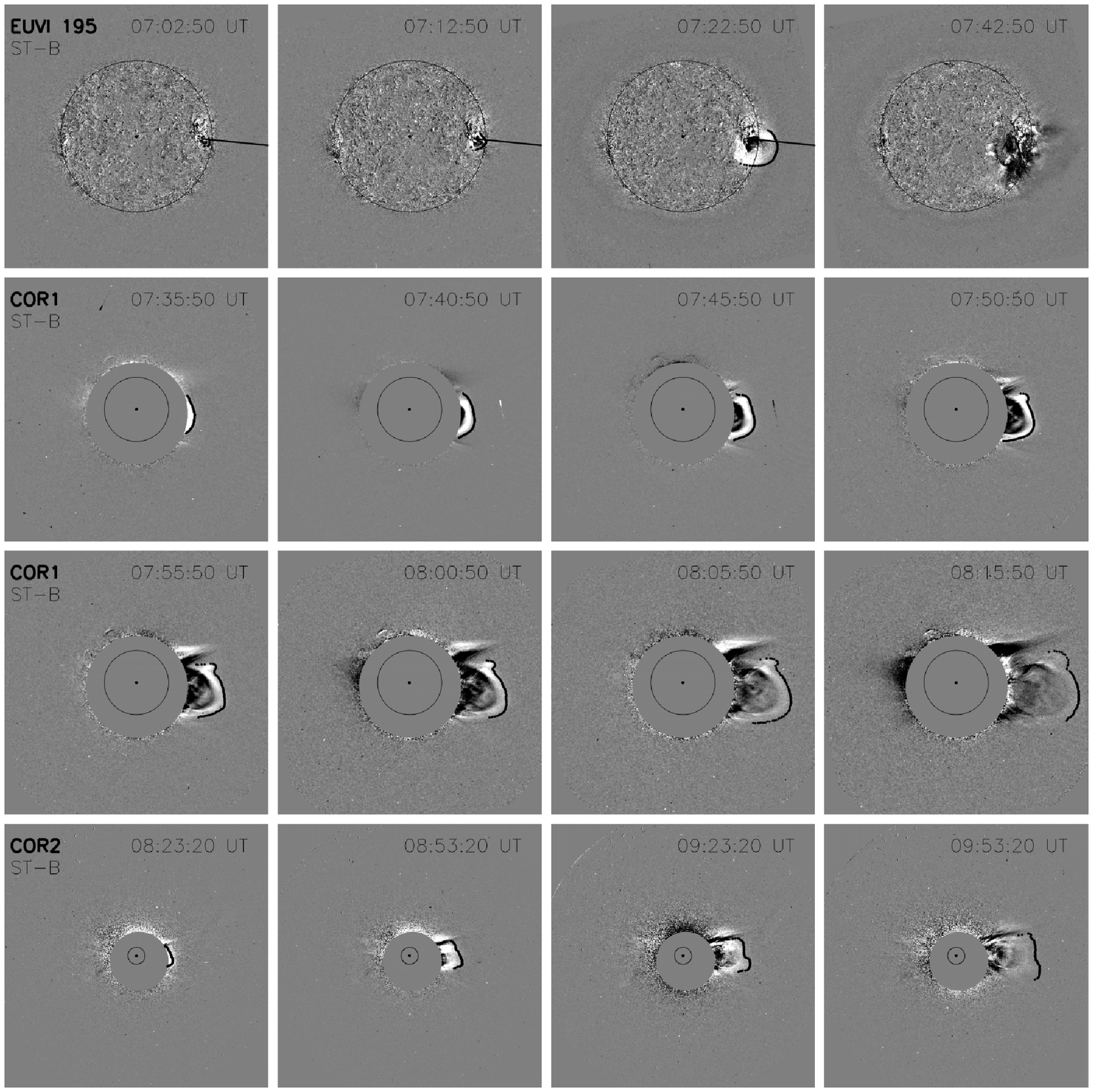}
		\caption{Same as Fig.~\ref{sequence} but for the CME event observed on 23 May 2007 by STEREO-B. The CME kinematics derived for this event is plotted in
		 Figs.~\ref{plotheight1} and \ref{plothva}. The ranges in $x$- and $y$-direction are [$-1580''$, $+1580''$] for EUVI 195~\AA, [$-3720''$, $+3720''$] for COR1 and 
		 $[-14\,000'',+14\,000'']$ for COR2.}
	\label{sequence1}
\end{figure*}

The Sun-Earth-Connection Coronal and Heliospheric Investigation (SECCHI) package \citep{howard2008} onboard the twin STEREO spacecraft, STEREO-A (Ahead) and STEREO-B (Behind), includes an Extreme Ultraviolet (EUV) Imager \citep[EUVI;][]{wuelser2004}, two white light coronagraphs (COR1, COR2) and two white light heliospheric imagers (HI1, HI2), observing the Sun in different FOVs. We combined data from EUVI, COR1 and COR2, in order to study the kinematics of CMEs from their initiation close to the solar surface up to a distance of about 15 $R_{\odot}$. EUVI observes the solar chromosphere and low corona in four different wavelengths in a FOV up to 1.7 $R_{\odot}$. To track a CME we mostly used images in the 171~\AA~passband because of the high time cadence up to 75 sec, but in some cases it was only possible to track the CME in the 195~\AA~observations, which on average have a lower time cadence (for most events 10 minutes, but in some cases as good as 2.5 minutes). The high time cadence enabled us to obtain detailed acceleration profiles especially at the beginning of the CME propagation. 

The two STEREO coronagraphs COR1 and COR2 observe the inner and outer solar corona in a FOV of 1.4 to 4 $R_{\odot}$ and 2.5 to 15~$R_{\odot}$, respectively. The overlapping FOVs of all three instruments enabled us to connect the same structure in the observations by the different instruments. The time cadence of the COR1 observations is mainly 5 minutes but can be up to 20 minutes in some cases, the cadence of COR2 is 30 minutes. 

We started our study with a sample of 146 selected CME events. For our selection we preferred events that could be observed already in the low corona 
(i.e. by the EUVI instrument), in order to get insights in the early phase of the CME dynamics. In addition, we required that the source region of a CME could be identified
on the visible solar hemisphere, for which we used the location of associated flares as an additional marker. Not each event fulfills both requirements (99 events could be measured in the EUVI FOV, for 89 events a flare could be associated). In most cases the associated flares were weak. Five events were associated with an M class flare, 26 CMEs with a C~class flare. The remaining flares were of GOES class B or below.\footnote{We note that the CME events in this study occurred during the extreme solar minimum period (2007--2010), which explains the lack of associated high-energy flare events.} 
The final CME sample for the statistics presented in this paper comprises a set of 95 events, for which we could derive the full CME acceleration profile, i.e. the peak acceleration and the acceleration duration could be measured for each CME. 84 out of them were tracked starting from the EUVI FOV.

All data were reduced by the SECCHI solar software routine \texttt{secchi$\_$prep}. The EUVI image sequences were corrected for differential rotation, and for weak events a normalizing-radial-graded filter \citep[NRGF;][]{morgan2006} was used. For the COR1 and COR2 observations a pre-event image was subtracted. In addition, a sigma filter and a normalization technique were applied to get better contrasts of the transient faint CME structures. For the measurements of the CME evolution, running difference and running ratio images were reconstructed.

\section{Methods and Analysis}
We derived the CME kinematics by measuring the position of the CME leading edge in EUVI, COR1 and COR2 images. In order to obtain reliable and reproducible measurements, an algorithm was developed which semi-automatically determines the position of the CME leading edge in EUVI, COR1 and COR2 running difference images. The leading edge of a CME appears as a bright front with a sharp intensity drop to regions outside the CME. This information is the basis for the algorithm developed to identify quasi-automatically the CME leading edge and its evolution in subsequent images. The algorithm works in the following way: The running difference images are contoured with brightness levels starting at very low, positive intensities. Then the distances of the pixels on the outermost contour with respect to Sun center are calculated. This process is repeated with contours at incrementally increasing brightness levels until the mean distance derived from two subsequent contour levels are separated by a sufficiently small, prescribed distance of each other. The position of the last determined contour is supposed to provide the location of the CME leading edge. The result can be manually corrected if necessary (e.g., if image artifacts are included in the contour, or if the CME front is not well defined). We note that faint CMEs were mostly measured by visual identification of the CME leading edge, since here the algorithm fails. Figures \ref{sequence} and \ref{sequence1} show examples of EUVI, COR1 and COR2 running difference image sequences of two well observed CMEs, where the identified CME leading edge and the determined propagation direction are indicated. We assumed a propagation along a straight line and defined for each event a direction, which crosses the outermost part of the CME front. The CME distance was derived by following the evolution of the identified CME leading edge along this main propagation direction. In a few cases a linear propagation could not be assumed due to deflection towards the heliospheric current sheet. For these events, the outermost part of the measured CME leading edge was manually selected in each frame.
The distance was averaged over an angular extent of $10^\circ$ of the determined leading edge and measured from the CME source region, which allows us to follow the CME propagation even if it deviates from radial direction. 

There are several factors affecting the determination of the CME leading edges and several possible sources for systematic errors. On the one hand, the CME itself can change its appearance with time, e.g., the front might become blurred which makes it difficult to track the same feature over several solar radii. On the other hand, we combine observations of different instruments with different angular resolution and the detection sensitivity varies over the FOV. In addition, stray light levels are different for COR1 and COR2, influencing the appearance of the observed white light feature. In order to estimate the average error included in our kinematical measurements, we analysed some selected CMEs in detail. For each of these events we followed four times the tracked CME leading edge in time in EUVI, COR1 and COR2 images. For each run, the scaling of the images was chosen differently, in order to account for the different visibilities of the outer (fainter) CME features. From the thus obtained height-time measurements, we derived the mean and standard deviation at each instant from the different measurement runs. The results from this procedure suggest an average error of 0.03 $R_{\odot}$ for measurements in EUVI, 0.125 $R_{\odot}$ in COR1, and 0.3 $R_{\odot}$ in COR2 data, respectively. 

Numerical differentiation of the height-time curve provides the CME velocity and acceleration profiles as a function of time. Due to the fact that errors in the height-time curve are enhanced by the first and second derivative, a smoothing and fitting method is used. We used a spline fitting procedure in which the measured CME height-time profile is subdivided into consecutive segments. All segments are then fitted by cubic splines and at their end points (`nodes'), the functions merge continuously and are twice continuously differentiable. The users's input to the fitting procedure is the number and position of nodes. The spline-fitted curve is then used as the basis to derive the velocity and acceleration evolution by subsequent numerical differentiation. In the course of fitting, we also estimated errors for the velocity and acceleration for each of the fitted segments. For each segment, the uncertainties on the polynomial spline coefficients are derived. The errors on the velocity and acceleration were then determined via Gaussian error propagation.

The advantage of using this method is that it provides reasonable errors (especially for the acceleration) also in cases of a high time cadence, where the classical error estimates using two neighbouring points yield typically very high errors for the CME accelerations. Different sets of nodes giving similar fits for a certain height-time curve may reveal considerably different velocity and acceleration profiles \citep[for details see Appendix in][]{vrsnak2007}. A fit was preferred over the other if the errors in velocity and acceleration were smaller, which basically confined the number of nodes to 5 to 7.

To estimate the CME source region size we used different methods, described in detail in \citet{vrsnak2007}:
\begin{itemize}
\item the distance between bipolar coronal dimming regions derived from EUVI 195~\AA~or 171~\AA~observations;
\item the footpoint separation of the associated eruptive filament/prominence observed in EUVI 304~\AA~images;
\item the lengths of the flare ribbon brightenings observed in the chromosphere or transition region measured in EUVI 304, 171 or 195~\AA~images.
\end{itemize}

\begin{figure}
	\centering
	\includegraphics[width=0.45\textwidth]{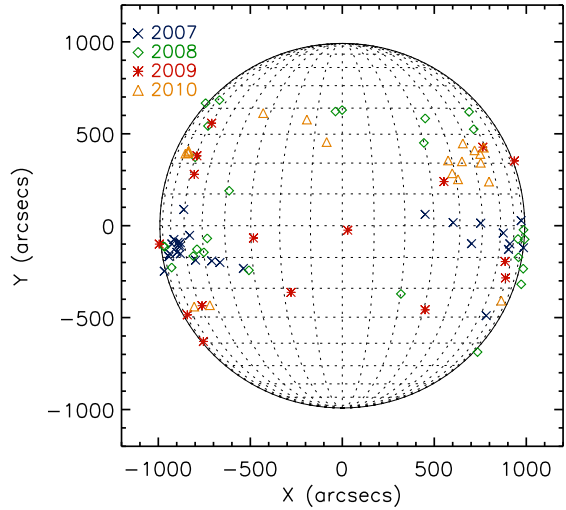}
		\caption{Positions of the CME source regions on the solar disc as observed from the actual STEREO vantage points. Different colors and symbols represent the different years of occurrence during the minimum phase of solar cycle no. 23/24.}
\label{sourcepos}
\end{figure}

\begin{figure}
	\centering
	\includegraphics[scale=1]{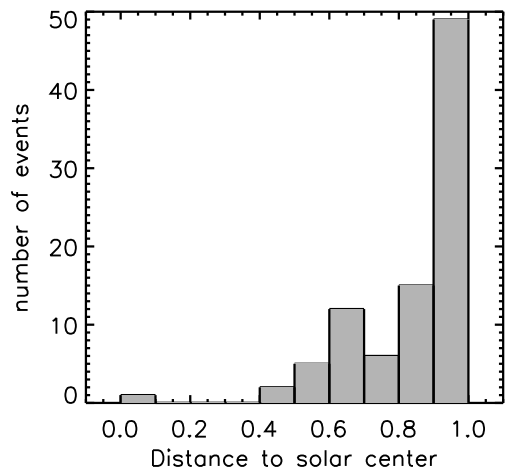}
		\caption{Distribution of the projected radial distances of the CME source regions from Sun center in units of $R_{\odot}$ with a binsize of 0.1~$R_{\odot}$.}
\label{histd}
\end{figure}

\section{Results}

\begin{figure*}
	\centering
	\includegraphics[scale=0.8]{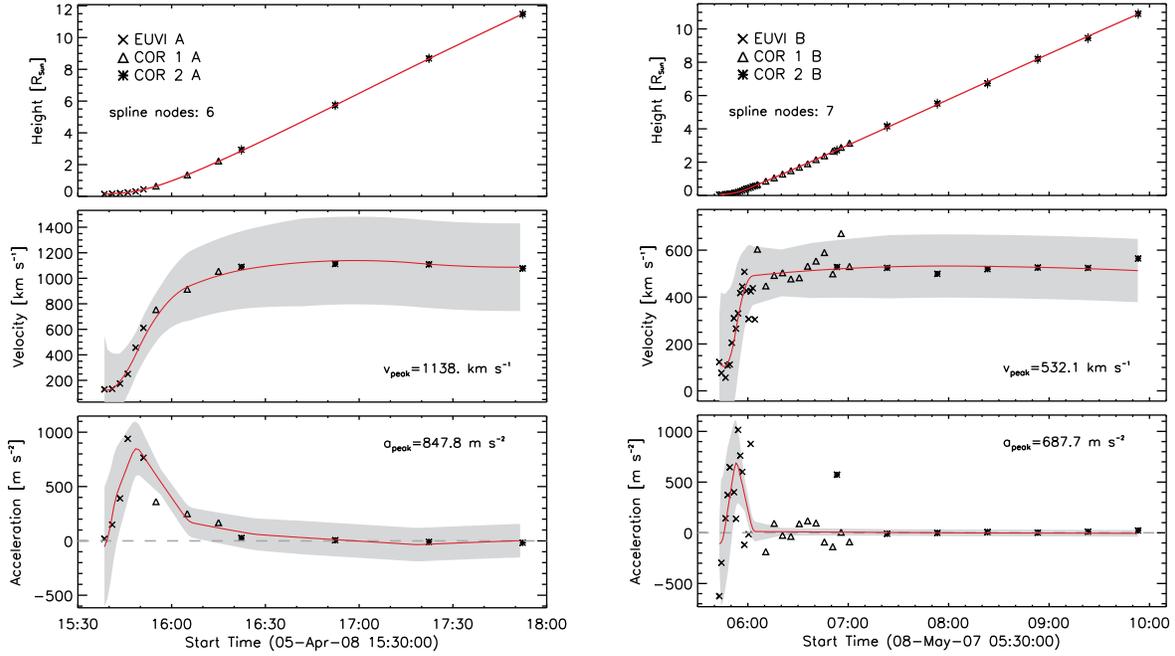}
		\caption{CME kinematics for the events observed on 5 April 2008 (left) and 8 May 2007 (right). The top panels show the height-time curve derived from EUVI (crosses), COR1 (triangles) and COR2 (asterisks) measurements together with the measurement errors. Note that these errors (0.03~$R_{\odot}$ for EUVI, 0.125~$R_{\odot}$ for COR1, and 0.3 $R_{\odot}$ for COR2 measurements) may appear smaller than the plot symbols due to the large height range presented. The solid line represents the spline fit to the height-time curve. The middle and bottom panels show the CME velocity and acceleration profiles derived from numerical differentiation of the CME height time measurements of the spline fit (solid line). The velocity and acceleration values derived by direct numerical differentiation of the measurements points (symbols) as well as the error range 
derived from the spline fits (grey shaded area) are overplotted. }
	\label{plotheight}
\end{figure*}

\begin{figure*}
	\centering
	\includegraphics[scale=0.8]{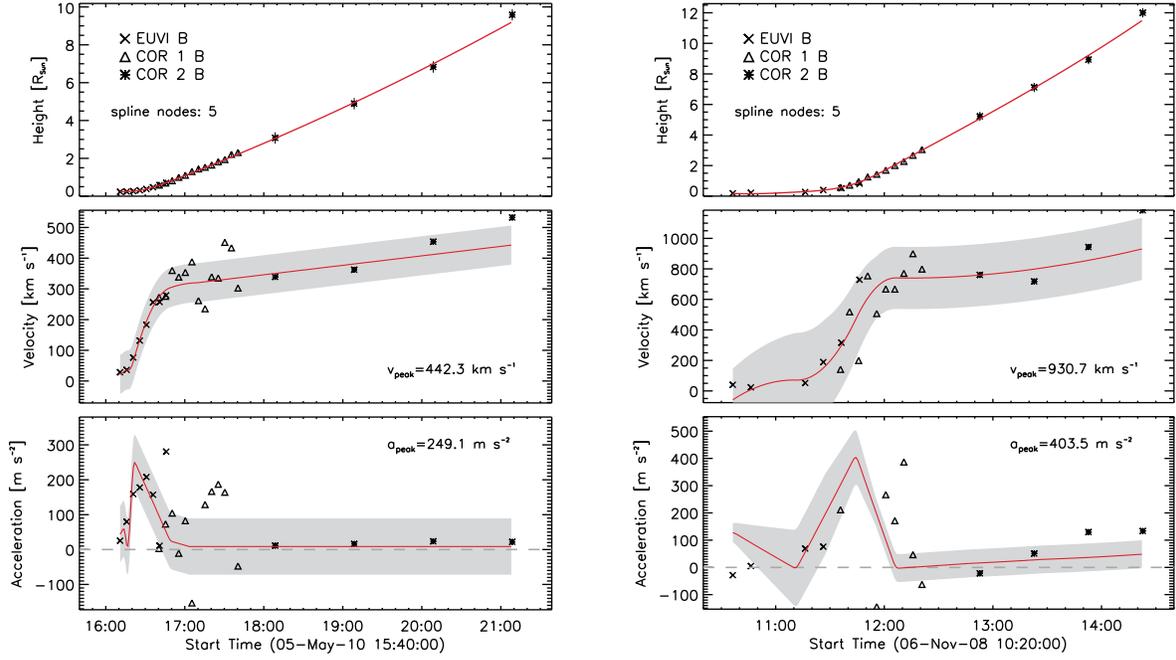}
		\caption{Same as Fig.~\ref{plotheight} but for the events observed on 5 May 2010 (left) and 6 November 2008 (right).}
	\label{plotheight1}
\end{figure*}

\begin{figure*}
	\centering
	\includegraphics[scale=0.8]{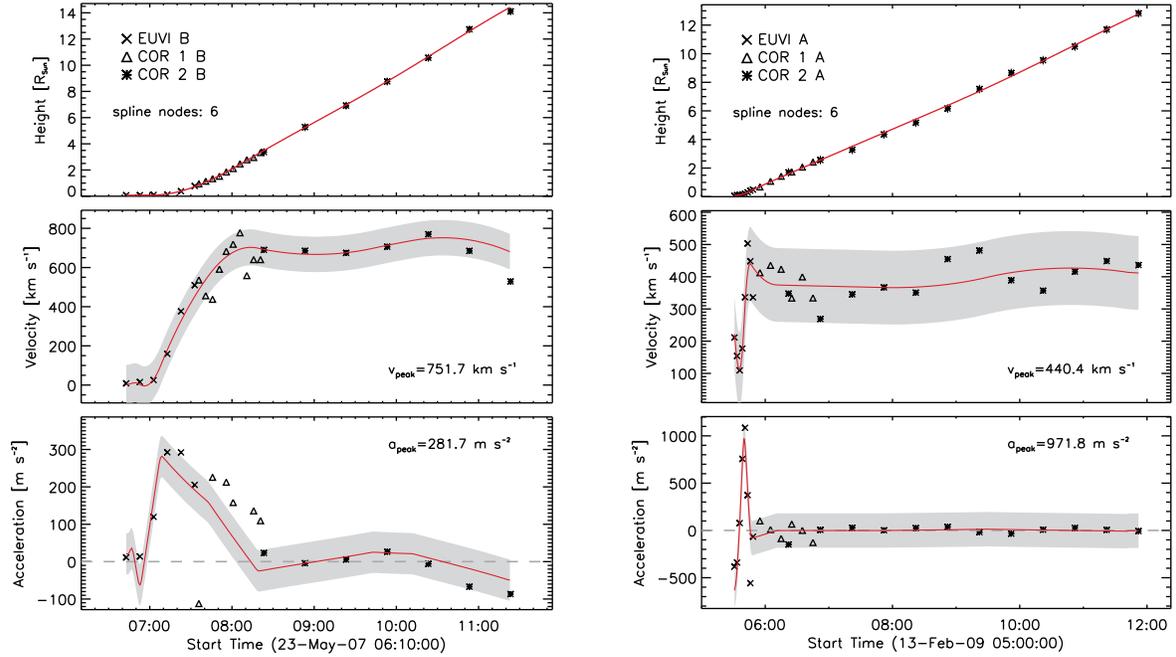}
		\caption{Same as Fig.~\ref{plotheight} but for the events observed on 23 May 2007 and 13 February 2009. }
	\label{plotheight2}
\end{figure*}

\begin{figure*}
	\centering
	\includegraphics[scale=0.75]{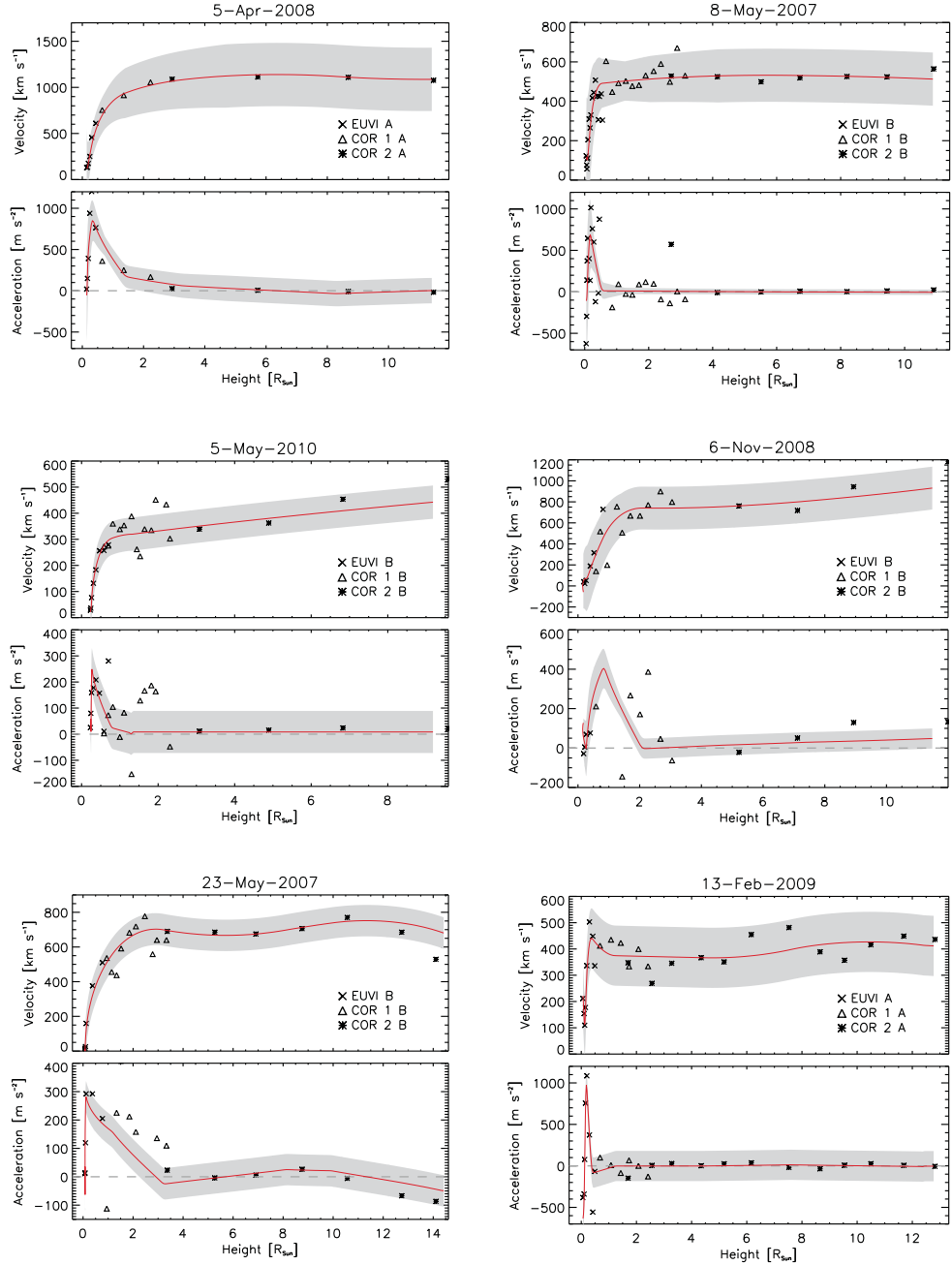}
		\caption{Evolution of the CME velocity and acceleration in dependence of the CME height for the events occurring on 5 April 2008, 8 May 2007 (top), 5 May 2010, 6 November 2008 (middle), 23 May 2007 and 13 February 2009 (bottom). Note that the height-time curves of these events are shown in Figs. \ref{plotheight}--\ref{plotheight2}. The data points derived from the different instruments are marked as crosses (EUVI), triangles (COR1) and asterisks (COR2). The solid lines indicate the first and second derivative of the spline fits, respectively, which are surrounded by the estimated error range (grey area). }
	\label{plothva}
\end{figure*}

\begin{figure}
	\centering
	\includegraphics[scale=1]{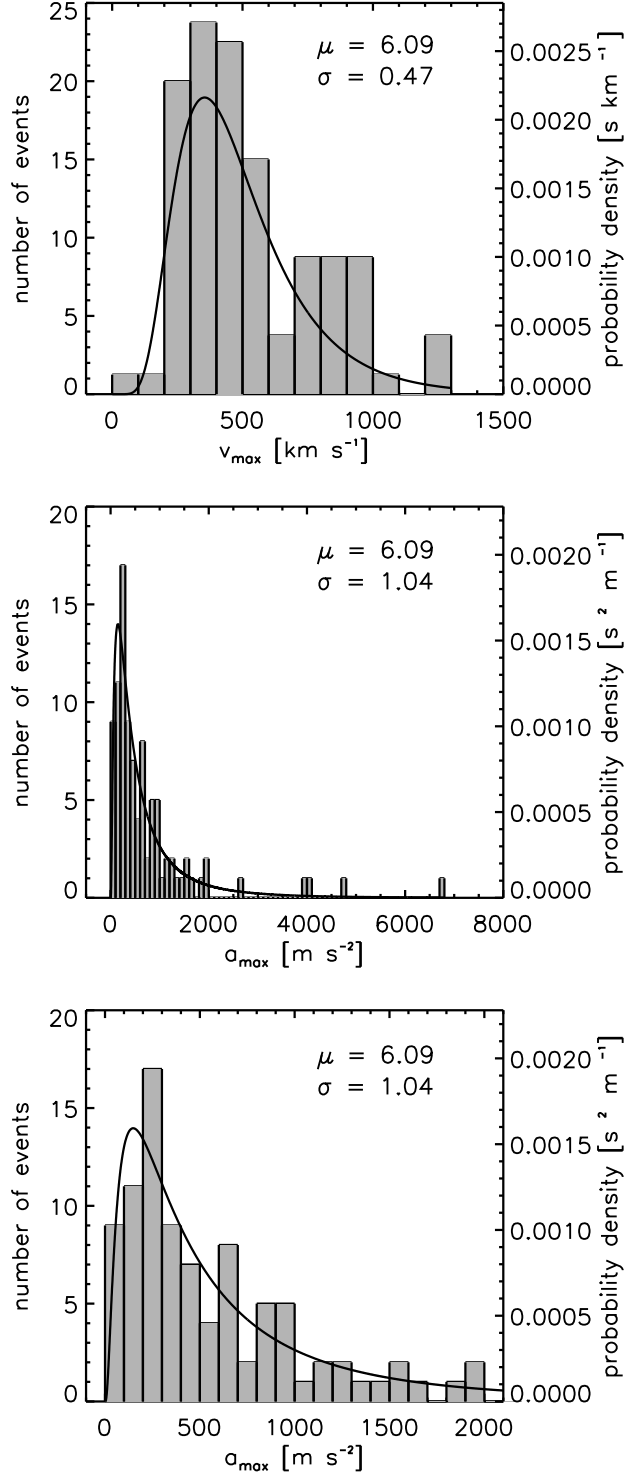}
		\caption{Distribution of the CME peak velocity (top) and peak acceleration (middle) derived for the whole sample of 95 events. The bottom plot shows a zoom of the middle plot restricted to peak accelerations $\leq$ 2000 m~s$^{-2}$. The solid line represents the lognormal fit with $\mu$ the mean and $\sigma$ the standard deviation, for details see main text.}
\label{histvel}
\end{figure}

\begin{figure}
	\centering
	\includegraphics[scale=1]{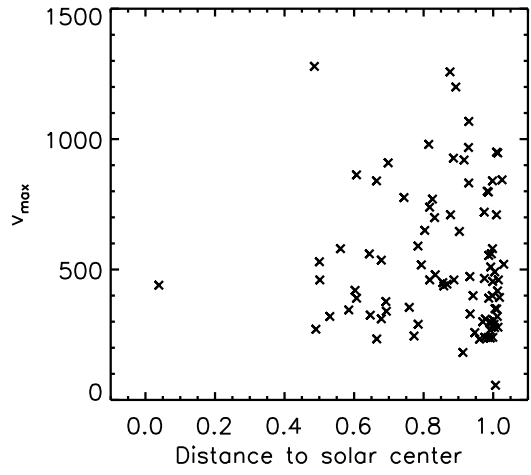}
		\caption{CME peak velocities $v_{max}$ against the projected radial distance to Sun center in units of $R_{\odot}$. }
\label{dist}
\end{figure}

\begin{figure}
	\centering
	\includegraphics[scale=1]{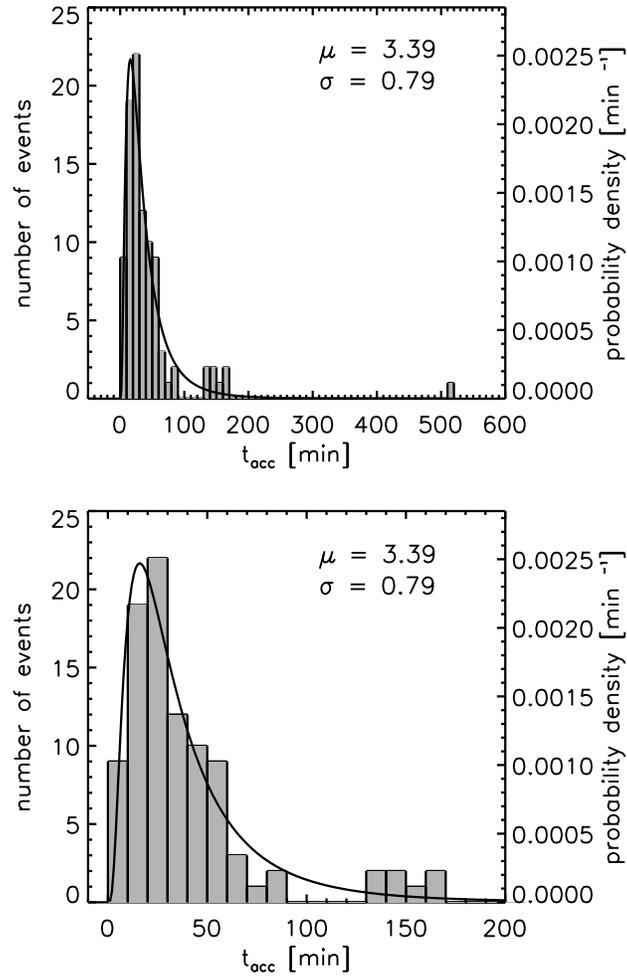}
		\caption{Distribution of CME acceleration duration $t_{acc}$ together with a lognormal fit. The bottom plot shows a zoom-in restricted to $t_{acc}<200$ min.}
\label{histacctime}
\end{figure}

\begin{figure}
	\centering
	\includegraphics[scale=1]{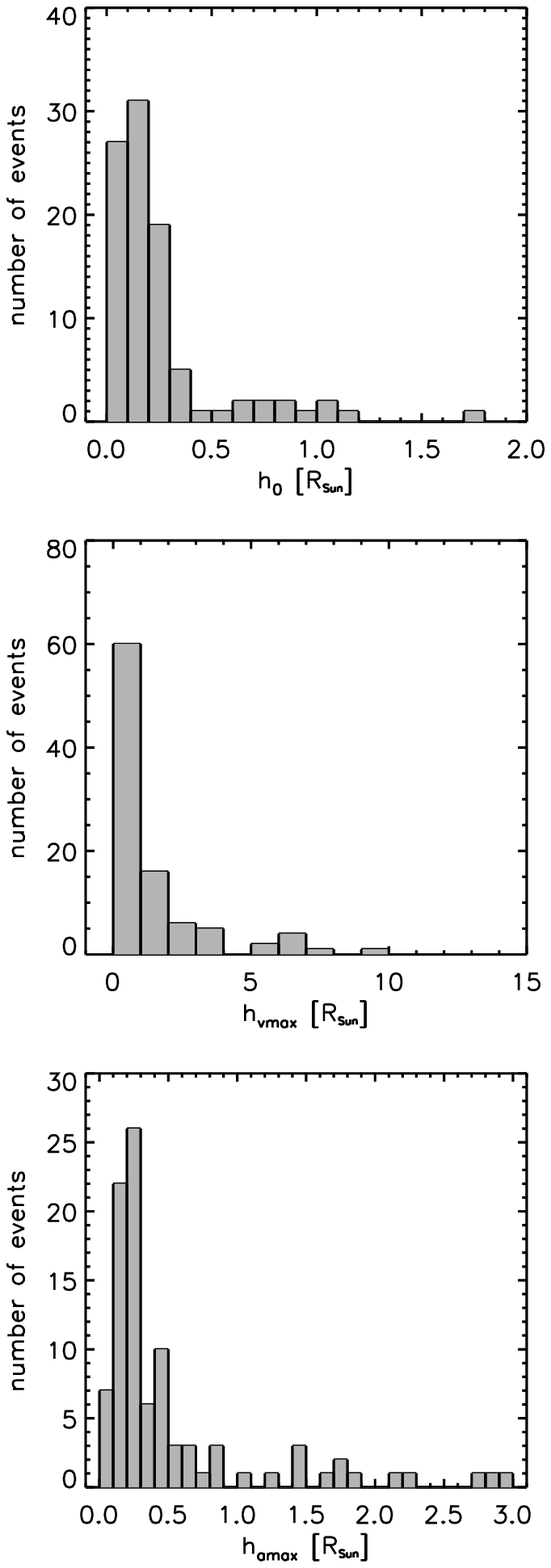}
		\caption{Distribution of $h_0$, i.e. the height at which the CME front could be first identified (top), height at peak velocity $h_{vmax}$ (middle) and height at peak acceleration $h_{amax}$ (bottom).}
\label{histh}
\end{figure}

In this paper a statistical analysis of 95 impulsive CME events is presented. For 90 events out of them a source region could be uniquely determined. Fig.~\ref{sourcepos} shows their position on the solar disc. Most of the events under study occurred close to the solar limb, and thus the influence of projection effects is small. Fig.~\ref{sourcepos} also reveals the transition from solar cycle no. 23 to no. 24, with the CME source regions changing from locations close to the equator at the end of cycle no. 23 to higher latitudes for CMEs already belonging to cycle no.~24. Fig.~\ref{histd} shows the distribution of the projected radial distance of the CME source regions to Sun center in units of $R_{\odot}$. A clear peak at 0.9 to 1 $R_{\odot}$ can be seen, which includes 49 events (more than 51\%). Only 21\% have distances smaller than 0.7~$R_{\odot}$, and only one event occurred close to disc center.

In Figs. \ref{sequence} and \ref{sequence1} we show EUVI and COR image sequences of two sample CMEs from our data set. The two events occurred on 5 April 2008 and 23 May 2007, respectively, and could be well observed in all three instruments, EUVI, COR1 and COR2. In both cases 16 representative images are selected; the whole evolution of the events can be seen in the accompanying movies no.~1 and no.~2. In each running difference image the solar limb is overplotted and the identified CME leading edges are marked. We measured the CME propagation along a straight line (also plotted into the image) which originates at the CME source region. In both examples the observed CMEs change their shape during their propagation until they reach the end of the COR2 FOV. Selected images from the different instruments, which observed the CME almost simultaneously, were overlaid to check if we really observe the same structure. This was possible for example for the EUVI image taken at 15:53:30 UT and the COR1 image taken at 15:55:00 UT in Fig.~\ref{sequence}. 

The kinematical plots derived for the CMEs shown in Figs. \ref{sequence} and \ref{sequence1} as well as for four more CME events from our sample can be seen in Figs. \ref{plotheight}--\ref{plotheight2}. The upper panels show the CME height, the middle panel the velocity and the bottom panel the acceleration evolution in time.  
Although all six events show the typical three phase kinematical behaviour (gradual initiation, acceleration and propagation phase), their kinematics reveal distinct differences. The event observed on 13 February 2009 (Fig.~\ref{plotheight2}) shows a very strong ($a_{max}$ $\sim$970 m~s$^{-2}$) and short acceleration phase of 9 min, whereas the events that occurred on 6 November 2008 (Fig.~\ref{plotheight1}) and 23 May 2007 (Fig.~\ref{plotheight2}) exhibit long acceleration durations of 50 and 72 min, respectively with peak accelerations of $\sim$400~m~s$^{-2}$ and $\sim$280~m~s$^{-2}$. For the event observed on 5 May 2010 (Fig.~\ref{plotheight}) an acceleration of 29~min and the smallest peak acceleration ($a_{max}$ $\sim$250~m~s$^{-2}$) of the six events shown in Figs. \ref{plotheight}--\ref{plotheight2}, was measured. The remaining two events plotted in Fig.~\ref{plotheight1} (5 April 2008, 8 May 2007) reveal peak accelerations of 850~m~s$^{-2}$ and 690~m~s$^{-2}$ and acceleration durations of 26~min and 17~min, respectively.

Fig.~\ref{plothva} shows for the six events plotted in Figs. \ref{plotheight}--\ref{plotheight2} the CME velocity and acceleration profiles against height, respectively, together with the error ranges derived from the spline fits. 
Differences are remarkable, in particular at which height the peak acceleration takes place. The event observed on 6 November 2008 reaches its maximum acceleration at a height of 0.84~$R_{\odot}$ above the CME source region, whereas for the other five events the height at peak acceleration is distinctly lower. For instance the event observed on 23 May 2007 was just at a height of 0.12~$R_{\odot}$ above the CME source region when it reached its peak acceleration of 280~m~s$^{-2}$. 

For each CME we derived several characteristic parameters:
\begin{itemize}
\item peak velocity $v_{max}$
\item peak acceleration $a_{max}$
\item acceleration duration $t_{acc}$
\item height at peak velocity $h_{vmax}$
\item height at peak acceleration $h_{amax}$
\item first height measured $h_0$
\item source region size $L$
\end{itemize}

The peak velocity $v_{max}$ and peak acceleration $a_{max}$ were derived from the spline fit of the velocity-time curve and the acceleration-time curve, respectively (see middle and bottom panels of Figs. \ref{plotheight}--\ref{plotheight2}). The measured velocity at the end of the COR2 FOV ($\sim$15~$R_{\odot}$) could result from the combination of two different effects, the main and the residual acceleration. If the residual acceleration is positive, the CME velocity profile peaks later but if the residual acceleration is negative, the maximum velocity is reached at the end of the main acceleration phase. The residual acceleration can be positive for two reasons. On the one hand, in slow and gradual events the ambient solar wind flow causes a further acceleration. On the other hand it is also possible that the CME accelerates continuously due to continuous energy release after the main phase \citep[e.g.][]{cheng2010}. Since we are interested in the CME velocity corresponding to the main acceleration phase, we determined $v_{max}$ as that value of the CME velocity, when the CME acceleration has decreased to 10\% of its peak value.

The acceleration duration $t_{acc}$=$t_{acc\_end}-t_{acc\_start}$ was extracted from the acceleration profile, where $t_{acc\_start}$ and $t_{acc\_end}$ were definded as the times when the CME acceleration profile is again at the 10\% level of the peak value. The height at peak velocity $h_{vmax}$ and height at peak acceleration $h_{amax}$ were derived from the velocity and acceleration profiles against the height (see Fig.~\ref{plothva}), $h_{vmax}$ being related to the 10\% level of $a_{max}$. The height $h_0$ at which the CME was first observed can be understood as a rough estimate of the height of CME initiation, since we observed all CMEs from their origin in the low corona. 

The top panel of Fig.~\ref{histvel} shows the distribution of the CME peak velocities. We found a range for $v_{max}$ from 56
to 1279~km~s$^{-1}$ with a mean value of 526~km~s$^{-1}$ and a median of 460~km~s$^{-1}$. The mode of the distribution lies at 300--400~km~s$^{-1}$. In comparison, \citet{vrsnak2007} measured 22 CMEs, which occurred in a period between February 2002 and January 2005, i.e. the maximum and decay phase of solar cycle no. 23, covering a range of 365 to 2775~km~s$^{-1}$ with a mean value of 940~km~s$^{-1}$. These linear velocity values are higher than the CME peak velocities in our study, which covers only events that occurred during the extreme solar minimum. Indeed, \citet{gopalswamy2009}, who derived the linear velocity for about 11000 LASCO CMEs that occurred between 1996 and 2006, found that the mean velocity values vary from 300 to 600~km~s$^{-1}$ during the solar cycle with an average value of 475~km~s$^{-1}$ for the whole sample. 

The distribution of the CME peak velocities $v_{max}$ derived in our study is asymmetrical with a tail towards high velocity values. This coincides with the findings of \citet{yur2005} for the distribution of the linear velocity of 4315 CMEs, which was fitted by a lognormal fit. The lognormal probability density function derived from our sample of $v_{max}$ is overplotted in Fig.~\ref{histvel} (top panel) as a solid line.
Mathematically, an independent variable $x$ is lognormally distributed when its natural logarithm, ln($x$), matches a normal distribution. A normal distribution is created by the sum of independent variables, whereas a lognormal distribution is created by the product of independent variables. In other words, if a variable is lognormally distributed, this hints at a multiplication of independent physical processes underlying the distribution \citep{limbert2001, yur2005}. The probability function $f(x)$ of the lognormal distribution can be written as:
\begin{equation}
f(x)=\frac{1}{\sqrt{2\pi}\sigma x}\exp\left( - \frac{(\ln(x)-\mu)^2}{2 \sigma^2}\right)
\end{equation}
where $\mu$ is the mean and $\sigma$ the standard deviation of the natural logarithm of $x$. With reference to \citet{limbert2001} we term $\mu^\ast=e^{\mu}$ and $\sigma^\ast=e^{\sigma}$ as the median and the multiplicative standard deviation. Thus the confidence interval of 68.3\% is given as [$\mu^\ast/\sigma^\ast, \mu^\ast\cdot\sigma^\ast$]. From the lognormal fit to the $v_{max}$ distribution (Fig.~\ref{histvel}, top panel) we obtained $\mu$=6.09 and $\sigma$=0.47 corresponding to a median $\mu^\ast$=441~km~s$^{-1}$ with the bounds of the confidence interval at [276~km~s$^{-1}$, 706~km~s$^{-1}$].

To check in which way projection effects influence our results we plot in Fig.~\ref{dist} the CME peak velocity against the radial distance of the CME source region to Sun center. The correlation coefficient of these two parameters is $c=$0.011, i.e. very low and thus projection effects do not have a significant effect. This is probably due to the fact that there exists a distinct correlation between the CME radial propagation and its lateral expansion \citep{schwenn2005}. Thus, the effect of projection effect on the derived CME velocities is much less than it could be expected in the case of point sources. 

The middle plot of Fig.~\ref{histvel} shows the distribution of the CME peak acceleration $a_{max}$ for the whole sample. The $a_{max}$ values cover a wide range between 19~m~s$^{-2}$ and 6781~m~s$^{-2}$. 40\% of the events have a peak acceleration $>$600~m~s$^{-2}$, 19\% show values $>$1000~m~s$^{-2}$ and five events (observed on 30 May 2007, 3 Jun 2007 $\sim$ 06:00 UT, 3 Jun 2007 $\sim$ 09:00 UT, 17 Aug 2007, 12 Feb 2010) reached $a_{max}$ values $>$2000~m~s$^{-2}$. The distribution peak is well-defined at 200--300~m~s$^{-2}$, the mean and median values are 756~m~s$^{-2}$ and 414~m~s$^{-2}$, respectively. For the distribution of the CME peak acceleration we again applied a lognormal fit, with the parameters $\mu$=6.09 and $\sigma$=1.04 corresponding to a confidence interval of [156~m~s$^{-2}$, 1248~m~s$^{-2}$] around $\mu^\ast$=441~m~s$^{-2}$. Since the majority of the $a_{max}$ values are concentrated in the range between 0 and 2000~m~s$^{-2}$, we plotted the detail of the histogram for that range at the bottom of Fig.~\ref{histvel}. 

The large range of $a_{max}$ values (19--6781~m~s$^{-2}$) spreading over two orders of magnitude is similar to former studies from \citet{vrsnak2007}, 40--7300~m~s$^{-2}$, and \citet{zhang2006}, 2.8--4464.9~m~s$^{-2}$. The mean value of $a_{max} = 749$~m~s$^{-2}$ in our study is comparable to the value from \citet{vrsnak2007}, 840~m~s$^{-2}$, but more than twice as high as the mean value in the sample of \citet{zhang2006}, 330~m~s$^{-2}$. It is worth noting that on average the CME peak velocities derived from CMEs that occurred during the extreme solar minimum (present study) are considerably smaller than those during solar maximum studied by \citet{vrsnak2007} but the CME peak accelerations are similar in both samples. This may be an effect of the better time cadence of the STEREO instruments which enables us to reconstruct fast changes in the CME kinematics.

The distribution of the acceleration phase duration $t_{acc}$, which peaks at 20--30~min, is shown in Fig.~\ref{histacctime} together with the lognormal fit. More than 50\% of the events have $t_{acc}$ values smaller than 30~min, however we measured also acceleration durations up to 8.6 hours. Only one event (observed on 17 October 2008) has an acceleration duration $>$200 min. Thus we show in the bottom panel of Fig.~\ref{histacctime} a zoom into the range of 0--200~min. The smallest value we obtained is 4.5~min. The arithmetic mean of 44.6~min is considerably lower than the values derived in \citet{vrsnak2007}, 120~min, and \citet{zhang2006}, 180~min, which indicates that our sample contains mainly impulsive events. The wide range of $t_{acc}$ of three orders of magnitude was also found by \citet{vrsnak2007} and \citet{zhang2006}. In both of these studies the distribution peaks at 0--50~min, consistent with our results. The distribution of $t_{acc}$ was fitted with the lognormal fit parameters $\mu$=3.39 and $\sigma$=0.79 which corresponds to a confidence interval of [13.5 min, 65.4 min] around $\mu^\ast$=29.7~min.

Fig.~\ref{histh} shows the distributions for the various CME height parameters, $h_0$, $h_{vmax}$ and $h_{amax}$. The distribution for $h_0$, the height at which the CMEs were first observed, is shown in the top panel of Fig.~\ref{histh}.
This is a measure for the height above the solar surface at which the CME is initiated. However, we stress that $h_0$ is not exactly the real CME initiation height but a rough measure for it, since it is affected by a) projection effects, and b)~sensitivity issues, i.e. the measured $h_0$ is expected to be larger than the real CME initiation height in cases where the CME could not be identified from the very beginning. The measured $h_0$ distribution covers the range from 0.01 to 1.76~$R_{\odot}$ with a mean value of 0.24~$R_{\odot}$ and a median value of 0.14~$R_{\odot}$. The maximum of the distribution is located at the low end at 0.1 to 0.2~$R_{\odot}$. The large $h_0$ values ($>0.6~R_{\odot}$) are from events which could not be identified in the EUVI FOV but only in coronagraphic images. These may be due to events that really start from source heights $\geq0.6~R_{\odot}$ or may be related to observational restrictions in terms of sensitivity for faint CMEs. For our sample of 95 events this applies to 11 CMEs.

The middle panel in Fig.~\ref{histh} shows the distribution of the heights $h_{vmax}$ defined as the velocity values reached at the end of the CME main acceleration phase. The distribution starts at very low heights of 0.17~$R_{\odot}$ and extends up to 9.5~$R_{\odot}$ (i.e. close to the border of the COR2 FOV). 63\% of the events are observed in the range 0--1~$R_{\odot}$. The mean and median values for $h_{vmax}$ are 1.46~$R_{\odot}$ and 0.78~$R_{\odot}$, respectively.

The bottom panel in Fig.~\ref{histh} shows the distribution of the heights $h_{amax}$ at which the CME accelerations reach their maximum. The mean of the $h_{amax}$ distribution is 0.53~$R_{\odot}$, the median 0.26~$R_{\odot}$. The peak in the $h_{amax}$ distribution lies between 0.2 and 0.3~$R_{\odot}$, and 74\% of the events have $h_{amax}$ smaller than 0.5~$R_{\odot}$. This means that most of the CMEs under study reach the acceleration at very low heights above the solar surface and emphasizes the importance of CME observations in the low corona in order to study the main acceleration phase.

Table 1 gives an overview of all the statistical CME parameters derived: we list the minimum and maximum value, the arithmetic mean with the standard deviation, the median together with the mean absolute deviation (mad) and the two fit parameters of the lognormal fit ($\mu$ and $\sigma$) for $v_{max}$, $a_{max}$, $t_{acc}$, $h_{0}$, $h_{amax}$ and $h_{vmax}$.

\begin{table*}
	\centering
		\begin{tabular}{||c | c | c| c |c | c ||}
		\hline
		\hline
		 &&&&&\\
			& mimimum & maximum & arithmetic mean $\pm$   & median $\pm$ mad & $\mu$  $\pm$ $\sigma$ \\
			&  &  &    standard deviation &      &  \\
			&&&&&\\
			\hline
			&&&&&\\
			$v_{max}$ [km~s$^{-1}]$ & 56 & 1279 & 526 $\pm$  263 & 460 $\pm$  160 & 6.09 $\pm$ 0.47 \\
			$a_{max}$ [m~s$^{-2}]$ & 19 & 6781 & 757 $\pm$  1034 & 414 $\pm$  246 & 6.09 $\pm$ 1.04\\
			$t_{acc}$ [min] & 4.5 & 516 & 44.6 $\pm$ 60.4 &  29.0 $\pm$ 14.5 &  3.39 $\pm$ 0.79 \\
			$h_{0}$ [$R_{\odot}$] & 0.01 & 1.76 & 0.24 $\pm$  0.29 & 0.14 $\pm$  0.08 & $-$  \\
			$h_{vmax}$ [$R_{\odot}$] & 0.17 & 9.5 & 1.56 $\pm$  1.82 & 0.78 $\pm$  0.42 & $-$  \\
			$h_{amax}$ [$R_{\odot}$] & 0.04 & 2.90 & 0.53 $\pm$  0.64 & 0.26 $\pm$  0.12 & $-$  \\
			&&&&&\\
		 \hline
		 \hline
		\end{tabular}
		\caption{We list the statistical CME parameters derived: the peak velocity $v_{max}$, the peak acceleration $a_{max}$, the acceleration phase duration $t_{acc}$, the height $h_{0}$, where the CME leading edge could be identified for the first time,  the height at peak velocity $h_{vmax}$, the height at peak acceleration $h_{amax}$. Minimum value, maximum value, arithmetic mean with standard deviation, median with the mean absolute deviation (mad) are derived from the whole data set of 95 events. $\mu$ and $\sigma$ are derived from the lognormal fit to the distribution.}
\end{table*}

\begin{figure}
	\centering
	\includegraphics[scale=0.95]{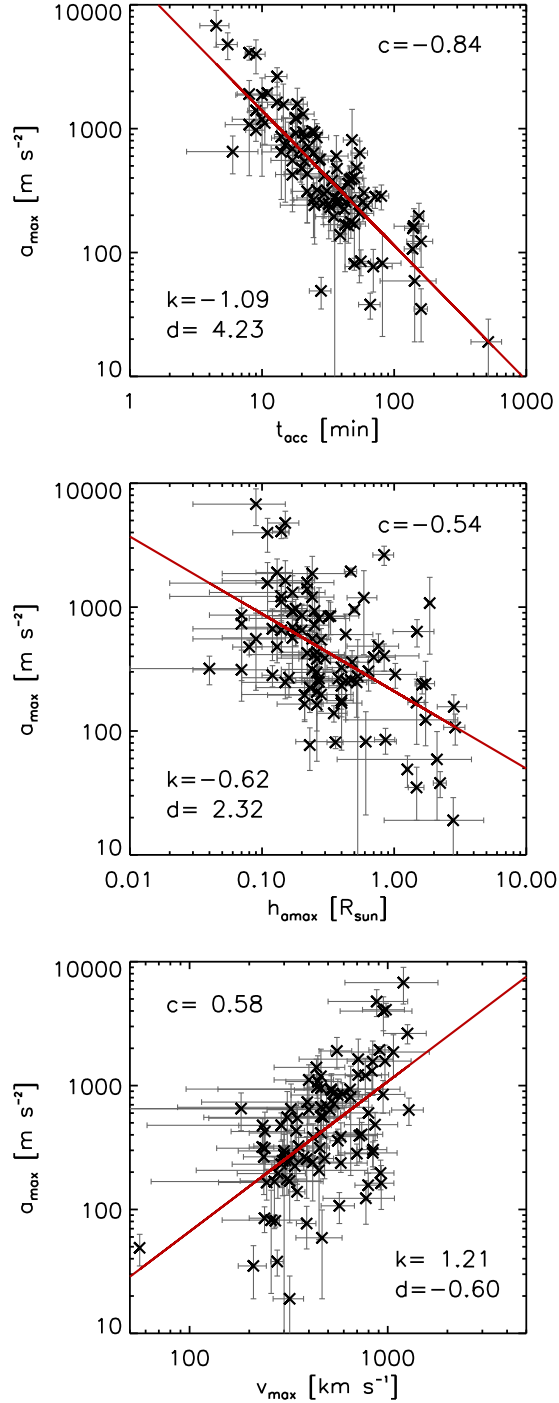}
		\caption{CME peak acceleration against acceleration duration (top), height at peak acceleration (middle) and peak velocity (bottom). The solid lines are linear regression lines to the data points with $k$ the slope and $d$ the $y$-intercept; $c$ gives the correlation coefficient. Note that $a_{max}$ as well as $t_{acc}$, $h_{amax}$ and $v_{max}$ are plotted on a logarithmic scale and that the fits and correlations were also determined in logarithmic space. The same holds for Figs. \ref{acctime} and \ref{h0}. }
\label{apeak}
\end{figure}

\begin{figure}
	\centering
	\includegraphics[scale=1]{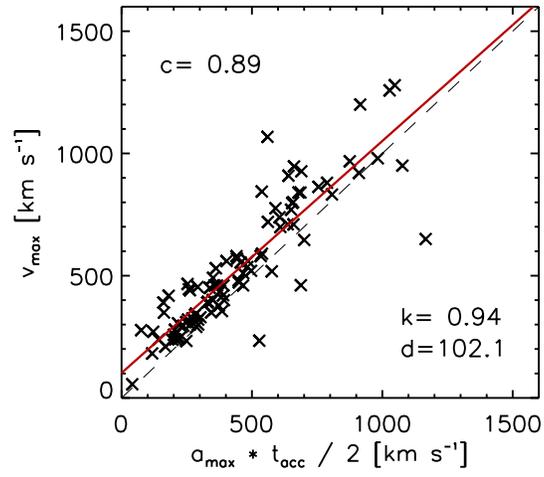}
		\caption{CME peak velocity against the acceleration duration multiplied by half of peak acceleration, together with the regression line (solid). For comparison also the 1:1 correspondence is plotted (dashed line).}
\label{vmax}
\end{figure}

\begin{figure}
	\centering
	\includegraphics[scale=1]{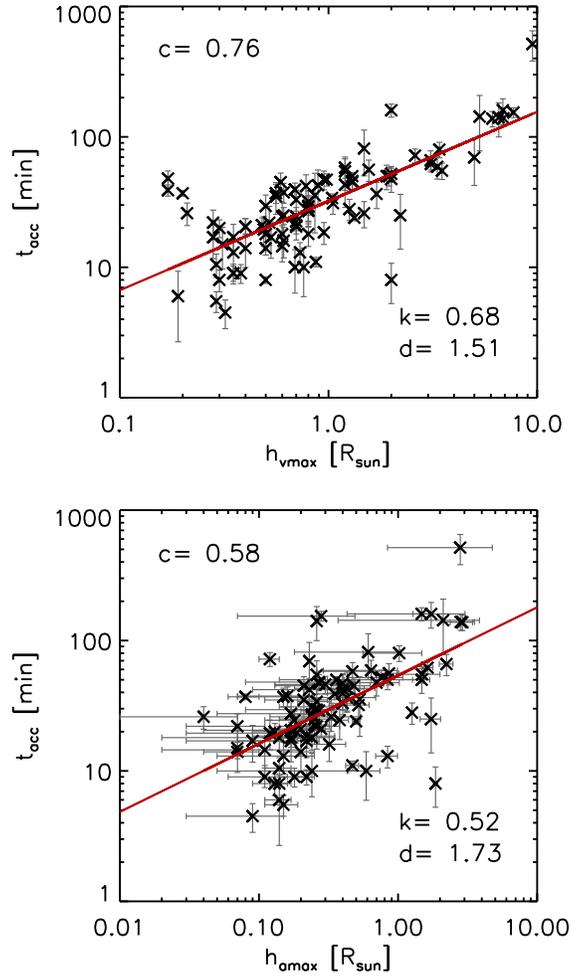}
		\caption{CME acceleration duration against height at peak velocity (top) and height at peak acceleration (bottom).}
\label{acctime}
\end{figure}

\begin{figure}
	\centering
	\includegraphics[scale=1]{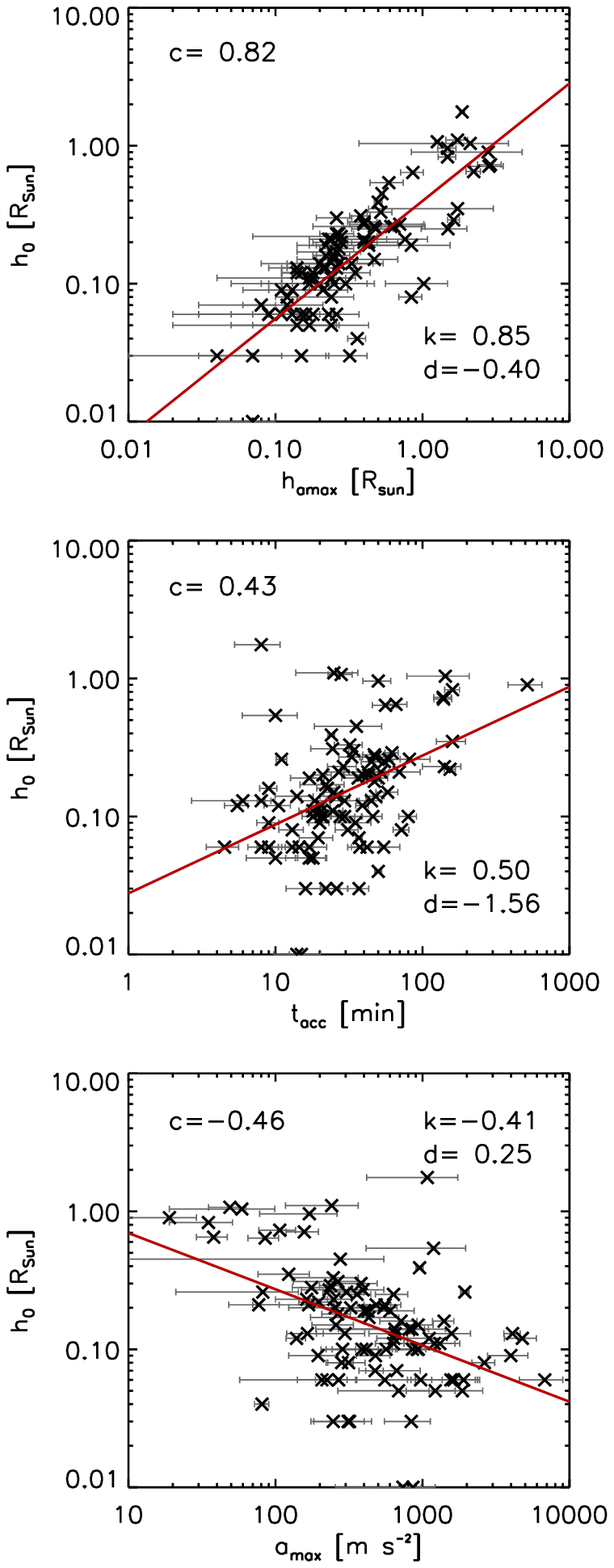}
		\caption{Height $h_{0}$ at which the CME front was first detected against the height at peak acceleration (top), acceleration duration (middle) and peak acceleration (bottom). }
\label{h0}
\end{figure}

In order to identify general characteristics and relationships, which intrinsically describe the evolution of CME eruptions, we correlated the various CME parameters derived. The correlation plots are shown in Figs. \ref{apeak}, \ref{acctime} and \ref{h0}. All the correlations are plotted and calculated in logarithmic space.
We found several CME parameters which revealed a significant correlation with the CME peak acceleration $a_{max}$. The top panel in Fig.~\ref{apeak} shows the scatter plot of $a_{max}$ against $t_{acc}$ revealing a high negative correlation with a correlation coefficient of $c=-0.84$, i.e. CMEs with larger peak accelerations have shorter acceleration durations. This finding fits well with the results obtained in former studies covering smaller CME samples \citep{zhang2006,vrsnak2007}. The slope ($-1.09$) of the regression line in Fig.~\ref{apeak} is in accordance with the results of \citet{vrsnak2007}, who found a slope of $-1.14$. The dependence between $a_{max}$ and $t_{acc}$ can be described with the following power law relation:\footnote{All relations between the different CME parameters are calculated in the same units as used in the scatter plots (Figs. \ref{apeak}, \ref{acctime} and \ref{h0}), i.e. m~s$^{-2}$ for $a_{max}$, km~s$^{-1}$ for $v_{max}$, min for $t_{acc}$ and $R_{\odot}$ for $h_{0}$, $h_{vmax}$ and $h_{amax}$.}
\begin{equation}
	a_{max}=10^{4.23}~t_{acc}^{-1.09}.
\end{equation}
The height $h_{amax}$ at which the CME reaches its maximum acceleration, and the peak acceleration $a_{max}$ are also anti-correlated, with $c=-0.54$ (Fig.~\ref{apeak}, middle), i.e. CMEs which are accelerated at lower heights reach higher peak accelerations. Between $a_{max}$ and $h_{amax}$ we found the following dependence:
\begin{equation}
	a_{max}=10^{2.32}~h_{amax}^{-0.60}.
\end{equation}
There is also a distinct correlation between $v_{max}$ and $a_{max}$, $c=0.58$ (Fig.~\ref{apeak}, bottom), which is not surprising, since CMEs which have stronger accelerations are also capable of reaching higher peak velocities. Between these two parameters we found the following power law dependence:
\begin{equation}
	a_{max}=10^{-0.60}~v_{max}^{1.21}.
\end{equation}
In contrast to the peak accelerations, the CME peak velocities do not show a significant correlation with $t_{acc}$, $h_{vmax}$ and $h_{amax}$ (plots not shown).

Fig.~\ref{vmax} shows as a test the CME peak velocity $v_{max}$ against $t_{acc}\, a_{max}/2$, which corresponds to modeling the CME acceleration by a triangular profile. 
We obtain a high correlation coefficient of $c=0.89$. For comparison, the regression line (solid line) is plotted together with the 1:1 correspondence (dashed line), which are nearly parallel but shifted against each other by $\sim$0--100~km~s$^{-1}$. This difference can be attributed to a residual acceleration of the CME not captured by the simple triangular profile assumed. For the linear regression line we found the following relation:
\begin{equation}
v_{max}=0.94~t_{acc}\frac{a_{max}}{2}+102.1,
\end{equation}
with the same units (km~s$^{-1}$) used on both sides of the equation.

The acceleration duration $t_{acc}$ against $h_{vmax}$ and $h_{amax}$ also shows a power law dependence with correlation coefficients of $c$=0.76 and $c$=0.58, respectively (Fig.~\ref{acctime}). Their power law dependence can be written as:
\begin{equation}
	t_{acc}=10^{1.51}~h_{vmax}^{0.68},
\end{equation}
\begin{equation}
	t_{acc}=10^{1.73}~h_{amax}^{0.52}.
\end{equation}
This means that CMEs which accelerate over a longer time also reach their maximum acceleration and velocity at larger heights.

In Fig.~\ref{h0} we show correlations of the height $h_{0}$, at which the CMEs were first observed against $h_{amax}$, $t_{acc}$ and $a_{max}$. A strong correlation between $h_{0}$ and $h_{amax}$ is found with a correlation coefficient of $c$=0.82, i.e.  CMEs which start at lower heights also reach their peak acceleration at lower heights. Between $h_{0}$ and $t_{acc}$ a weak correlation of $c$=0.43 is found, i.e. CMEs which originate at low heights tend to accelerate more impulsively.
Between $h_{0}$ and $a_{max}$ an anti-correlation with $c=-0.46$ was found, i.e. CMEs starting at lower heights in the solar corona reach larger peak accelerations. The relations can be expressed as:
\begin{equation}
	h_{0}=10^{-0.40}~h_{amax}^{0.85},
\end{equation}
\begin{equation}
	h_{0}=10^{-1.56}~t_{acc}^{0.50},
\end{equation}
\begin{equation}
	h_{0}=10^{0.25}~a_{max}^{-0.41}.
\end{equation}

\begin{figure}
	\centering
	\includegraphics[scale=1]{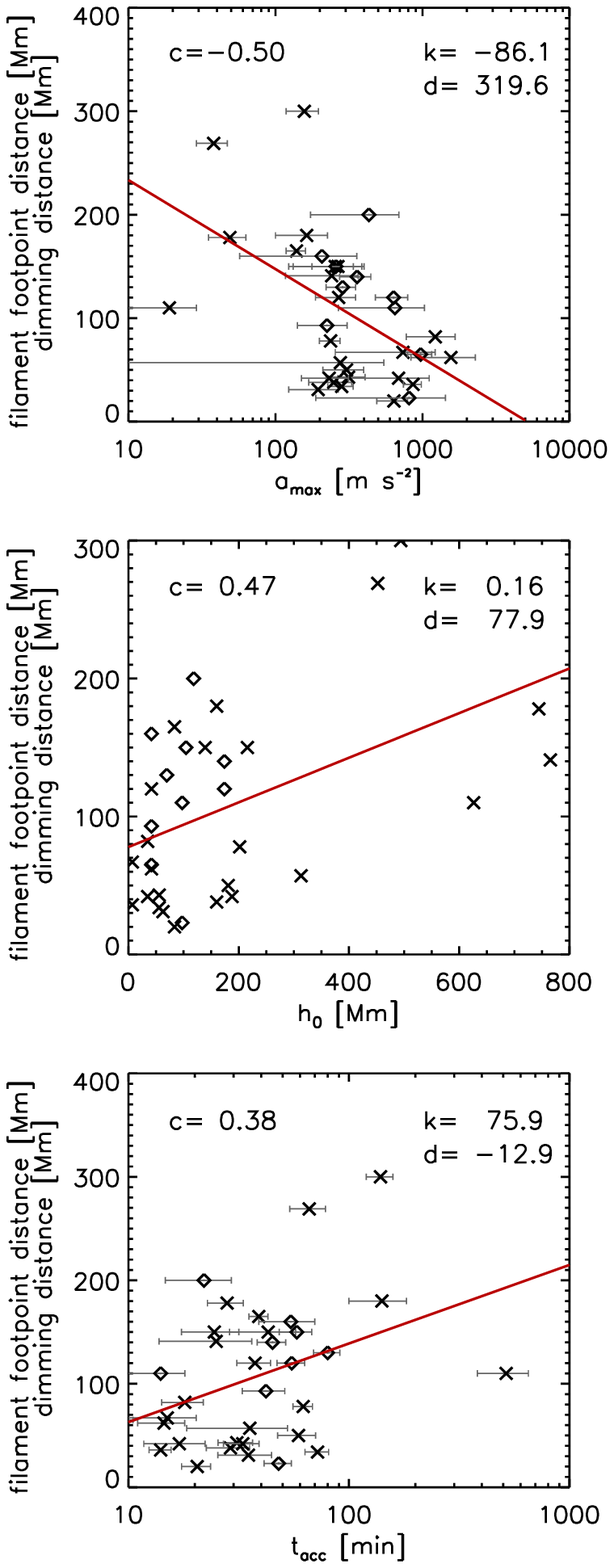}
		\caption{Estimates of the CME source region size against the CME peak acceleration (top), the height of the first CME observation (middle) and acceleration duration (bottom). Crosses indicate the distance of the associated filament footpoints, diamonds the spatial extent of the coronal dimming region.}
\label{source}
\end{figure}

For 78 out of the 95 events it was possible to measure at least one feature to estimate the CME source region size. Because there are general differences between the different measurement methods we considered each feature separately. The flare ribbon length of the associated flares did not show any distinct correlation with the CME peak acceleration and velocity but the footpoint distances of the erupting filaments and the size of the coronal dimming do.
Fig.~\ref{source} shows the filament footpoint distances (measured for 24 events) together with the linear extent of the associated coronal dimmings (measured for 10 events) against $a_{max}$, $h_0$ and $t_{acc}$.
The negative correlation between the source region size $L$ and $a_{max}$ ($c=-$0.50) and the positive correlation between $L$ and $t_{acc}$ ($c=$0.38) indicates again that CMEs which originate from compact sources reach higher peak accelerations and have shorter acceleration durations consistent with the findings of \citet{vrsnak2007}.
Between the CME source region size $L$ and initiation height $h_0$ a positive correlation was found with the correlation coefficient of $c=$0.47.

\section{Summary and Conclusions}

In the following we summarize the most important findings of our study based on a sample of 95 impulsive CMEs observed in STEREO EUVI, COR1 and COR2.
\begin{enumerate}
\item The histograms of the CME peak velocity $v_{max}$, the CME peak acceleration $a_{max}$ and the CME acceleration duration $t_{acc}$ can be approximated with a lognormal distribution.
\item We found a wide range of values for $a_{max}$ (19--6781~m s$^{-2}$) and $t_{acc}$ (4.5~min--8.6~h).
\item Most of the events (74\%) reach their peak acceleration at heights $<$0.5 $R_{\odot}$.
\item $a_{max}$ and $t_{acc}$ are negatively correlated, $c=-$0.84.
\item $a_{max}$ and the height at peak acceleration $h_{amax}$ are negatively correlated, $c=-$0.54.
\item $t_{acc}$ and the height at peak velocity $h_{vmax}$ are positively correlated, $c=$0.76.
\item $t_{acc}$ and $h_{amax}$ are positively correlated, $c=$0.58.
\item $t_{acc}$ and the CME initiation height $h_{0}$ are positively correlated, $c=$0.43.
\item $h_{0}$ and $h_{amax}$ are positively correlated, $c=$0.82.
\item $h_{0}$ and $a_{max}$ are negatively correlated, $c=-$0.46.
\item The CME source region size $L$ and $a_{max}$ are negatively correlated, $c=-$0.50.
\item $L$ and $h_{0}$ are positively correlated, $c=$0.47.
\item $L$ and $t_{acc}$ are positively correlated, $c=$0.38.
\end{enumerate}

Based on the assumption that the Lorentz force is the main driver of the eruption, it can be assumed that magnetic energy is transformed into kinetic energy and thus
\begin{equation}
\frac{\rho v^2}{2}~\leq~\frac{B^2}{2\mu_0}
\end{equation}
or
\begin{equation}
v~<~\left(\frac{B^2}{\mu_0 \rho}\right)^{1/2}~=~v_A,
\label{vel}
\end{equation}
with $B$ the magnetic field strength within the CME body, $\mu_0$ the magnetic permeability in vacuum, $v_A$ the Alfv\'{e}n velocity, $\rho$ the plasma density and $v$ the CME velocity. Eq. \ref{vel} states that the CME velocity cannot be larger than the Alfv\'{e}n velocity in the erupting structure \citep[e.g.,][]{vrsnak2006}.

\begin{figure}
	\centering
	\includegraphics[scale=1]{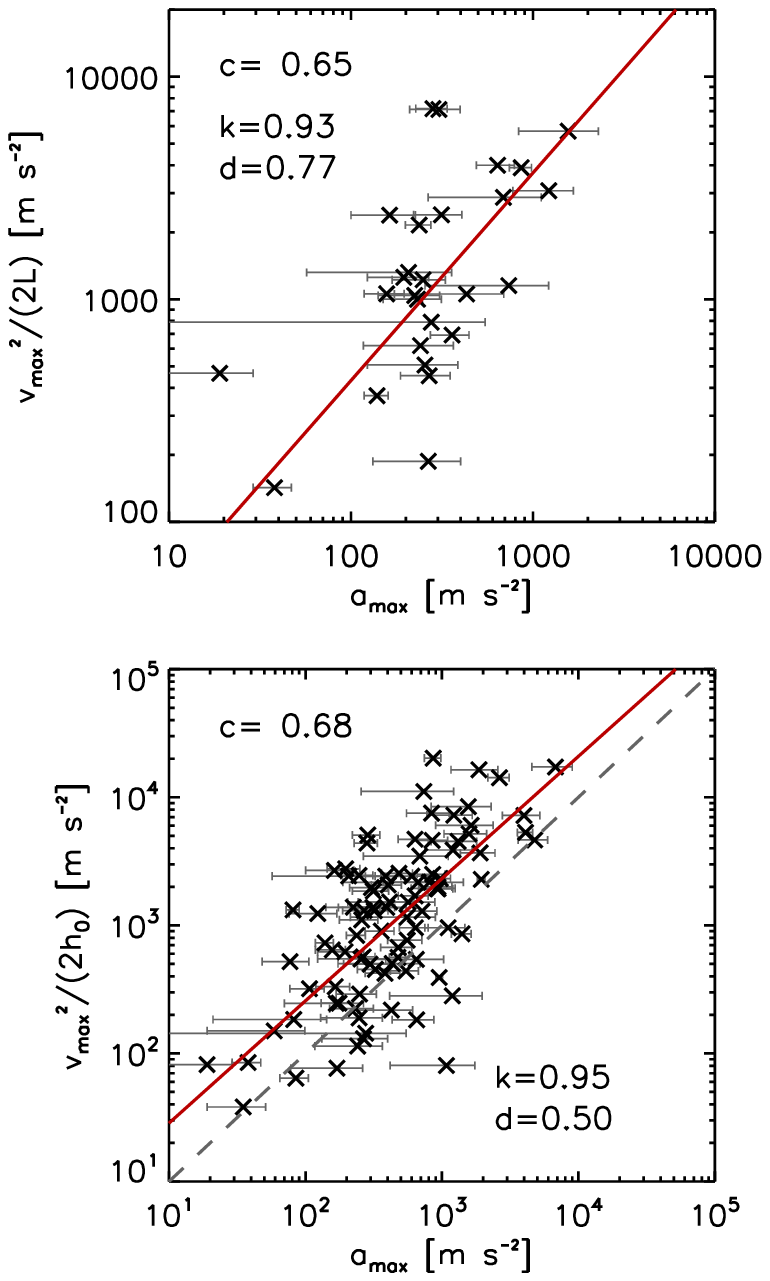}
		\caption{Top: Relation of Eq. \ref{acc} with the CME source region size $L$ (filament footpoint distance, dimming extent). Bottom: Same relation but with the height $h_0$ at which the CME was first observed instead of $L$. }
\label{v2d}
\end{figure}

The Lorentz force density can be expressed as:
\begin{equation}
\boldsymbol{f}=\boldsymbol{j}\times\boldsymbol{B}=
\frac{1}{\mu_0}\left[(\boldsymbol{B}\cdot\boldsymbol{\nabla})\boldsymbol{B}-\boldsymbol{\nabla}\left(\frac{\boldsymbol{B}^2}{2}\right)\right]
\end{equation}
with $\boldsymbol{j}$ the current density. Inserting the approximation $\boldsymbol{\nabla}\approx 1/L$, we obtain the following order of magnitude estimate \citep[see also][]{vrsnak2007}:
\begin{equation}
a~<~\frac{B^2}{2 \mu_0 \rho L}~=~\frac{v_A^2}{2L}
\label{acc}
\end{equation}
with $L$ the characteristic length scale over which the magnetic field varies, which can be approximated by the CME source region size.
Eq. \ref{acc} shows that the acceleration is not only governed by the Alfv\'{e}n velocity but is also dependent on the size of the erupting structure. Initially compact CMEs (small $L$, large $v_A$) will reach higher accelerations. These considerations coincide with our observational findings (summary item 10 and 11), i.e. the inverse proportionality found between $a_{max}$ and $L$ (Fig.~\ref{source}, top) and also $a_{max}$ and $h_0$ (Fig.~\ref{h0}, bottom), which can be used as an alternative estimate of the source region size. We tested  the  relation between the CME peak acceleration, the size of the erupting flux rope and the Alfv\'{e}n velocity implied by Eq.~\ref{acc}, in that we derived for each event the ratio $v_{max}^2/(2L)$ and $v_{max}^2/(2h_0)$ (with $v_{max} < v_A$, see Eq.~\ref{vel}), and correlated these quantities with $a_{max}$. The resulting correlation coefficients lie in the range of 0.6--0.7 (Fig.~\ref{v2d}), i.e. are higher than the correlations of the individual CME parameters, which supports our interpretation.

The distinct anticorrelation found between  $h_{0}$ and $a_{max}$ with $c \sim -0.5$ is in addition also related to the stronger magnetic fields in the lower corona, which in turn are related to larger Lorentz forces providing the driving force for the CME acceleration. We note that we interpret the parameter $h_0$, defined as the height where a CME is first observed, to be a measure of the initiation height of the erupting CME (and thus, to some extent, also to be a measure of the original size of the erupting structure).
This is of course somewhat critical, since in gradual and faint CMEs we expect that we cannot really observe a CME from its very inititiation site
(due to the limited sensitivity) but only further out when sufficient mass was accumulated at the CME front to be observable. Indeed, in 11 out of 95 events we could not find signatures of the erupting CME in the EUVI but only in the COR1 FOV. This may be a real effect, i.e. some CMEs in our sample started at heights $\gtrsim 0.7\,R_\odot$, 
but may be partly also biased by the fact that the CME actually started at lower heights but was to faint to be observed. However, even if we 
underestimated the CME initiation height by using the height $h_0$ where the CME was first observed, the distinct correlations 
that we obtained between $h_0$ and other characteristic CME quantities such as $a_{max}$, $v_{max}$ and $t_{acc}$, 
in line with the interpretations in terms of the Lorentz force elaborated above, support it to be a very useful quantity for CME initiation studies.

Since during its propagation the size of a CME increases with height, we expect from Eq.~\ref{acc} also that the acceleration decreases with height. 
This is consistent with our findings of a distinct anticorrelation between $a_{max}$ and $h_{amax}$ (summary item no. 5, see also bottom panel of Fig.~\ref{apeak}).
The distinct inverse proportionality that we derived between $a_{max}$ and $t_{acc}$ (summary item no. 4, see top panel of Fig.~\ref{apeak}) explains the relative small range of CME velocities (1 order of magnitude) despite the large ranges, over which $a_{max}$ (2 orders of magnitude) and $t_{acc}$ (3 orders of magnitude) vary.

Taking into account that the acceleration time cannot be shorter than the Alfv\'{e}n wave signal transit time, we get the order-of-magnitude relation:
\begin{equation}
t_{acc}~\geq~\frac{2L}{v_A},
\label{time}
\end{equation}
i.e., CMEs originating from compact sources accelerate more impulsively, consistent with our summary item no. 13 (Fig.~\ref{source}, bottom) and no. 8 (Fig.~\ref{h0}, top), where we interpret $h_0$ as another measure of the source region size.

The distributions of the CME peak velocity, CME peak acceleration and acceleration duration show a lognormal behavior (summary item no.1). Such distributions are created by the product of several independent variables. We suggest that the energy of the CME is dependent on at least two major variables. On the one hand, it depends on the amount of initially stored magnetic energy, which is available for transformation into other forms of energy. Furthermore, the CME energy depends also on the ``transmission coefficient", i.e., the percentage of the initially stored energy which will be transformed into kinetic energy of the CME.

\acknowledgements
This work was supported by the \"Osterreichische F\"orderungsgesellschaft (FFG) of the Austrian Space Applications Programme (ASAP) under grant no. 819664 and by the Austrian Science Fund (FWF): P20867-N16. The European Community's Seventh Framework Programme (FP7/2007-2013) under grant agreement no. 218816 (SOTERIA) is
acknowledged by B.V. We thank the anonymous referee for insightful comments, which helped to improve the paper. 
The STEREO/SECCHI data are produced by an international consortium of the Naval Research Laboratory (USA), Lockheed Martin Solar and Astrophysics Lab (USA), NASA Goddard Space Flight Center (USA), Rutherford Appleton Laboratory (UK), University of Birmingham (UK), Max-Planck-Institut f\"ur Sonnensystemforschung (Germany), Centre Spatiale de Li\`{e}ge (Belgium), Institut d'Optique Th\'{e}orique et Appliqu\'{e}e (France), and Institut d'Astrophysique Spatiale (France).



\end{document}